\documentstyle[aps,prl,twocolumn,pstcol]{revtex}
\begin{document}
\draft
\title{Nonlinear Perturbation Theory}
\author{St\'ephanie Rossano and Christian Brouder}
\address{Laboratoire de Min\'eralogie-Cristallographie, Universit\'es Paris 6 et 7, 
IPGP, case 115, 4 place Jussieu, 75252 Paris cedex 05, France}
\date{\today}
\maketitle
\begin{abstract}
An explicit perturbative solution to all orders 
is given for a general class of nonlinear differential
equations. This solution is written as a sum indexed by
rooted trees and uses the Green function of a 
linearization of the equations.
The modifications due to the presence of
zero-modes is considered. Possible divergence of
the integrals can be avoided by using approximate Green
functions.
\end{abstract}
\pacs{}

\newcommand{\ima}{{\mathrm{i}}}
\newcommand{\dd}{{\mathrm{d}}}
\newcommand{\sech}{{\mathrm{sech}}}
\newcommand{\phiperp}{\varphi_{\perp}}

\newcommand{\tun}{\setlength{\unitlength}{2pt}
\psset{unit=2pt}
\psset{runit=2pt}
\psset{linewidth=0.2}
\begin{pspicture}(0,0)(2,2)
\psdots[dotstyle=*](1,1)
\end{pspicture}}

\newcommand{\tdeux}{\setlength{\unitlength}{3pt}
\psset{unit=3pt}
\psset{runit=3pt}
\psset{linewidth=0.2}
\begin{pspicture}(0,0)(2,2)
\psline(1,0)(1,2)
\psdots[dotstyle=*](1,0)
\psdots[dotstyle=*](1,2)
\end{pspicture}}

\newcommand{\ttroisun}{\setlength{\unitlength}{3pt}
\psset{unit=3pt}
\psset{runit=3pt}
\psset{linewidth=0.2}
\begin{pspicture}(0,0)(2,2)
\psline(1,0)(0,2)
\psline(1,0)(2,2)
\psdots[dotstyle=*](1,0)
\psdots[dotstyle=*](0,2)
\psdots[dotstyle=*](2,2)
\end{pspicture}}

\newcommand{\ttroisdeux}{\setlength{\unitlength}{3pt}
\psset{unit=3pt}
\psset{runit=3pt}
\psset{linewidth=0.2}
\begin{pspicture}(0,0)(2,4)
\psline(1,0)(1,2)
\psline(1,2)(1,4)
\psdots[dotstyle=*](1,0)
\psdots[dotstyle=*](1,2)
\psdots[dotstyle=*](1,4)
\end{pspicture}}

\newcommand{\tquatreun}{\setlength{\unitlength}{3pt}
\psset{unit=3pt}
\psset{runit=3pt}
\psset{linewidth=0.2}
\begin{pspicture}(0,0)(2,4)
\psline(1,0)(0,2)
\psline(1,0)(2,2)
\psline(2,2)(2,4)
\psdots[dotstyle=*](1,0)
\psdots[dotstyle=*](0,2)
\psdots[dotstyle=*](2,2)
\psdots[dotstyle=*](2,4)
\end{pspicture}}

\newcommand{\tquatredeux}{\setlength{\unitlength}{3pt}
\psset{unit=3pt}
\psset{runit=3pt}
\psset{linewidth=0.2}
\begin{pspicture}(0,0)(2,4)
\psline(1,0)(0,2)
\psline(1,0)(2,2)
\psline(0,2)(0,4)
\psdots[dotstyle=*](1,0)
\psdots[dotstyle=*](0,2)
\psdots[dotstyle=*](2,2)
\psdots[dotstyle=*](0,4)
\end{pspicture}}

\newcommand{\tquatretrois}{\setlength{\unitlength}{3pt}
\psset{unit=3pt}
\psset{runit=3pt}
\psset{linewidth=0.2}
\begin{pspicture}(0,0)(2,4)
\psline(1,0)(1,2)
\psline(1,2)(2,4)
\psline(1,2)(0,4)
\psdots[dotstyle=*](1,0)
\psdots[dotstyle=*](1,2)
\psdots[dotstyle=*](2,4)
\psdots[dotstyle=*](0,4)
\end{pspicture}}

\newcommand{\tquatrequatre}{\setlength{\unitlength}{3pt}
\psset{unit=3pt}
\psset{runit=3pt}
\psset{linewidth=0.2}
\begin{pspicture}(0,0)(2,6)
\psline(1,0)(1,2)
\psline(1,2)(1,4)
\psline(1,4)(1,6)
\psdots[dotstyle=*](1,0)
\psdots[dotstyle=*](1,2)
\psdots[dotstyle=*](1,4)
\psdots[dotstyle=*](1,6)
\end{pspicture}}

\newcommand{\tquatrecinq}{\setlength{\unitlength}{3pt}
\psset{unit=3pt}
\psset{runit=3pt}
\psset{linewidth=0.2}
\begin{pspicture}(0,0)(4,2)
\psline(2,0)(0,2)
\psline(2,0)(2,2)
\psline(2,0)(4,2)
\psdots[dotstyle=*](2,0)
\psdots[dotstyle=*](0,2)
\psdots[dotstyle=*](2,2)
\psdots[dotstyle=*](4,2)
\end{pspicture}}


The Born expansion is a common tool of quantum mechanics.
It states that, for a Hamiltonian $H=H_0+V$, a solution of 
$H\varphi=E\varphi$ is given by 
$\varphi=\varphi_0+\sum_{n=0}^\infty (GV)^n\varphi_0$,
where $\varphi_0$ is a solution of $H_0\varphi_0=E\varphi_0$,
and $G=(E-H_0)^{-1}$ is the Green function corresponding to
the unperturbed problem. The Born expansion has rendered
innumerable services for pratical as well as theoretical
problems of quantum mechanics. A drawback of this expansion
is that it is restricted to linear problems.

In the present paper, the Born expansion will be extended
to non-linear problems. More precisely, we consider 
an equation of the type $F_0(\varphi)=F_1(\varphi)$,
where $F_0$ and $F_1$ are functionals of $\varphi$. 
$F_0$ describes the unperturbed system and $F_1$ 
its perturbation.
For instance, the propagation of ideal optical solitons
in an optical fiber is governed by the non-linear Schr\"odinger equation,
\cite{Essiambre,Hasegawa}
\begin{eqnarray}
F_0(\varphi)= \ima\frac{\partial \varphi}{\partial z} +
\frac{1}{2}\frac{\partial^2 \varphi}{\partial \tau^2}+
|\varphi|^2\varphi. \label{nlse}
\end{eqnarray}
We assume that we know a solution of the equation
$F_0(\varphi_0)=0$. For Eq.(\ref{nlse}), such a solution is
the optical soliton
\begin{eqnarray*}
\varphi_0(z,\tau)=\frac{\eta \exp\big(\ima (\eta^2-\xi^2)z/2 + \ima\xi\tau\big)}
{\cosh \big(\eta (\tau-\xi\zeta)\big)},
\end{eqnarray*}
where $\eta$ and $\xi$ are parameters.

In a true optical material, the ideal soliton is
perturbed by various effects represented by the 
perturbation \cite{Hasegawa}
\begin{eqnarray*}
F_1(\varphi)=-\ima\Gamma\varphi +\ima\beta 
\frac{\partial^3 \varphi}{\partial \tau^3} 
-\alpha_1 \frac{\partial (|\varphi|^2\varphi)}{\partial \tau}
+\alpha_2 \varphi \frac{\partial |\varphi|^2}{\partial \tau}.
\end{eqnarray*}

We will show that, if $F_0$ and $F_1$ are functionally differentiable,
an explicit expression
for the solution of $F_0(\varphi(x))=F_1(\varphi(x))$
is the Butcher series
\begin{eqnarray}
\varphi(x)=\varphi_0(x)+\sum_{t} \frac{1}{\sigma(t)}
\bar\varphi(t;x). \label{solution}
\end{eqnarray}

The first part of the paper will be devoted to the explanation 
and derivation of Eq.(\ref{solution}), then the role of 
zero modes will be investigated. Finally, we show how
approximate Green functions can be used to avoid
divergent integrals.

The basic tool for the derivation of Eq.(\ref{solution})
is that we sum over all rooted trees $t$. A tree is a graph
without loop, and a rooted tree is a tree where one vertex
is designated as its root. We draw the root at the bottom
of the tree.
The smallest rooted tree is the single root $\tun$.
The rooted trees with up to 4 vertices are
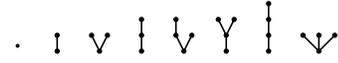
\begin{eqnarray*}
\tun\quad\tdeux\quad\ttroisun\quad\ttroisdeux\quad\tquatredeux
\quad\tquatretrois\quad\tquatrequatre\quad\tquatrecinq
\end{eqnarray*}

Rooted trees
have been introduced in 1857 by A. Cayley to represent
derivatives of a function with respect to a parameter
\cite{Cayley}.
In 1963, J.C. Butcher used rooted trees to write the
solution of flow equations and derive new numerical methods
\cite{Butcher63}. Since that work, series indexed by
rooted trees are called B-series or Butcher series.
In 1998 Kreimer discovered that rooted trees underlie
the basic mathematical structure of renormalization theory
\cite{Kreimer98}.

In Eq.(\ref{solution}), $|t|$ designates the number of
edges of $t$. For instance, $|\tdeux |=2$.
To define the symmetry factor $\sigma(t)$, we need the
collecting operator $B_+$,\cite{Connes} which starts from $k$ trees
$t_1,\dots,t_k$ and builds a new tree $t=B_+(t_1,\dots,t_k)$
by joining the root of each of the $k$ trees to a new
vertex that becomes the root of $t$. 
For instance $B_+(\tun)=\tdeux$,
$B_+(\tun,\tun)=\ttroisun$. The edges can be rotated around
the vertices, so that
\begin{eqnarray*}
B_+(\tun,\tdeux)=B_+(\tdeux,\tun)=\tquatredeux.
\end{eqnarray*}
Every rooted tree $t$ (except the root) can be written as
$t=B_+(t_1,\dots,t_k)$ for some $t_1,\dots,t_k$.
Finally, $\sigma(t)$ is the symmetry factor of tree $t$. It
is defined recursively by $\sigma(\tun)=1$,
\begin{eqnarray}
\sigma(B_+(t_1^{n_1} \dots t_k^{n_k})) &=&
n_1!\sigma(t_1)^{n_1}\dots n_k!\sigma(t_k)^{n_k}. \label{defsigma}
\end{eqnarray}
The notation $t=B_+(t_1^{n_1} \dots t_k^{n_k})$ means that
$t$ is obtained by collecting $n_1$ times tree
$t_1$,\dots, $n_k$ times tree $t_k$,
where the $k$ trees
 $t_1$, \dots, $t_k$ are all different.

The last term to define in Eq.(\ref{solution}) is $\bar\varphi(t;x)$.
We recall that the functional derivative of $F(\varphi(x))$ with
respect to a function $\psi(x)$ is \cite{Glimm}
\begin{eqnarray*}
\frac{\delta F(\varphi(x))}{\delta\psi(x)}=
\lim_{\epsilon\to 0}\frac{F(\varphi(x)+\epsilon\psi(x))-
F(\varphi(x))}{\epsilon}.
\end{eqnarray*}
This is called the G{\^a}teaux derivative in the mathematical
literature \cite{Balakrishnan}.
Then the functional derivative 
\begin{eqnarray*}
\frac{\delta F(\varphi(x))}{\delta\varphi(y)}
\,\,{\mathrm{is}}\,\,{\mathrm{defined}}\,\,{\mathrm{as}}\,\,
\frac{\delta F(\varphi(x))}{\delta\psi(x)}\,\,{\mathrm{for}}\,\,
\psi(x)=\delta(y-x).
\end{eqnarray*}

From $F_0$ and the unperturbed solution $\varphi_0(x)$ we
define the operator $M$ with kernel $M(x,y)$
\begin{eqnarray*}
M(x,y)=\frac{\delta F_0(\varphi_0(x))}{\delta\varphi(y)}.
\end{eqnarray*}
Now we define the corresponding Green function $G(x,y)$
by $\int \dd y M(x,y)G(y,z)=\delta(x-z)$. The definition
of $G(x,y)$ involves the boundary conditions imposed on 
$\varphi(x)$. A method to construct $G(x,y)$ for solitons
was proposed by  Kawata and Sakai \cite{Kawata}.

We can write $F_0(\varphi(x))$ as a sum over its functional
derivatives:
\begin{eqnarray*}
F_0(\varphi(x)) &=& F_0(\varphi_0(x))+ 
\int \dd y M(x,y)(\varphi(y)-\varphi_0(y)) \\ && +F_2(\varphi(x)),
\end{eqnarray*}
where
\begin{eqnarray*}
F_2(\varphi(x)) &=&
\sum_{n=2}^\infty \frac{1}{n!}
\int \dd y_1\dots\dd y_n \frac{\delta^n F_0(\varphi_0(x))}
{\delta\varphi(y_1)\dots\delta\varphi(y_n)}\\&&
\times
(\varphi(y_1)-\varphi_0(y_1))\dots(\varphi(y_n)-\varphi_0(y_n)).
\end{eqnarray*}
Since $F_0(\varphi_0(x))=0$, the equation 
$F_0(\varphi)=F_1(\varphi)$ can be rewritten
\begin{eqnarray*}
\int \dd y M(x,y)(\varphi(y)-\varphi_0(y))=F(\varphi(x)),
\end{eqnarray*}
where $F(\varphi(x))=F_1(\varphi(x))-F_2(\varphi(x))$.
We operate the last equation by the Green function $G$
to obtain
\begin{eqnarray*}
\varphi(x)=\varphi_0(x)+\int \dd y G(x,y)F(\varphi(y)).
\end{eqnarray*}
This equation has exactly the form of Eq.(25) in Ref.\cite{BrouderEPJC1},
and the proof given in Ref.\cite{BrouderEPJC1} can be followed to show
that $\varphi(x)$ is given by Eq.(\ref{solution}), where
$\bar\varphi(t;x)$ is defined recursively by
\begin{eqnarray*}
\bar\varphi(\tun;x) &=& \int \dd y G(x,y) F(\varphi_0(y))
\end{eqnarray*}
for the root and, for $t=B_+(t_1,\dots,t_k)$, by
\begin{eqnarray*}
\bar\varphi(t;x) &=& 
 \int \dd y \dd z_1 \dots \dd z_k G(x,y) 
\frac{\delta^k F(\varphi_0(y))}{\delta \varphi(z_1)\dots\delta
\varphi(z_k)}\\&&\times\bar\varphi(t_1;z_1)\dots\bar\varphi(t_k;z_k).
\end{eqnarray*}

To summarize, once the initial function $\varphi_0(x)$ and
the corresponding Green function $G(x,y)$ are known, the
calculation of $\varphi(x)$ up to any order 
is a mechanical application of a simple formula,
which is well suited to computer algebra programs.
To estimate the size of the terms in
Eq.(\ref{solution}), we can use the fact that,
if $F_1(\phi)$ is written as $\epsilon F_1(\phi)$,
then $\bar\phi(t;x)$ is a sum of terms ranging
from $\epsilon^{(|t|+1)/2}$ to $\epsilon^{|t|}$.

To be complete, we must investigate the influence of
zero modes. The term zero modes come from the theory of
instantons in quantum field theory \cite{Rajaraman,Zinn-Justin}.
Within our framework, they are the solutions $\psi_n(x)$ of the
equation $\int \dd y M(x,y)\psi_n(y)=0$. In other words, 
the zero modes are a basis of the kernel $K$ of the operator $M(x,z)$.

In a linear problem, the zero modes are the solution of the
unperturbed equation $H\varphi_0=E\varphi_0$, and the superposition
principle tells us that, if $\varphi$ (resp. $\varphi'$) is 
the solution given by the Born expansion starting from the zero mode
$\varphi_0$ (resp. $\varphi_0'$), then we can start from 
$\varphi_0+\varphi_0'$ to obtain another solution
($\varphi+\varphi'$). The proper $\varphi_0$ is determined
by the boundary conditions. 

For a nonlinear problem, the superposition principles 
does not hold. For notational convenience, we assume that
$\dim K=1$ and $\psi_0\in K$. We write the solution of
the nonlinear problem as
$\varphi=\varphi_0 + \phiperp +\lambda \psi_0$, 
where $\phiperp\in K^{\perp}$. 
As in the first part of the paper and in \cite{BrouderEPJC1},
we transform the original problem into an equation of the
form 
\begin{eqnarray}
\varphi(x)=\varphi_0(x)+F(\varphi(x)). \label{Butcherform}
\end{eqnarray}
Then, such equation can
be solved immediately using Butcher series, as was also
noticed indepently by Schatzman \cite{Schatzman} and
Connes and Kreimer \cite{ConnesK}. 
Notice that Butcher series can be obtained for
quite general $F(\varphi(x))$: $F$ can be a
function of $x$ (i.e. $F(\varphi(x),x))$, it can
involve differentials, $\varphi(x)$ and $x$ can
be multidimensional  \cite{BrouderEPJC1}.

Now, we start from the equation $F_0(\varphi)=F_1(\varphi)$,
and we propose a possible method to bring the
problem into the form (\ref{Butcherform}). 
From $\varphi=\varphi_0 + \phiperp +\lambda \psi_0$ we can write
\begin{eqnarray*}
F_0(\varphi) &=&
 F_0(\varphi_0)+M\phiperp+\lambda M \psi_0+F_2(\varphi),\\
F_1(\varphi) &=&
 F_1(\varphi_0)+\lambda \frac{\delta F_1(\varphi_0)}{\delta \psi_0}+
  F_3(\varphi),
\end{eqnarray*}
where 
$F_2$ and $F_3$ are defined by the equations and
$\delta F_1(\varphi_0(x))/\delta \psi_0(x)=\int \dd y \psi_0(y)
\delta F_1(\varphi_0(x))/\delta \varphi(y)$ and
$M(x,y)=\delta F_0(\varphi_0(x))/\delta \varphi(y)$.

Since $F_0(\varphi_0)=0$ and  $M\psi_0=0$, we obtain
\begin{eqnarray*}
M\phiperp-\lambda \frac{\delta F_1(\varphi_0)}{\delta \psi_0}=
F_1(\varphi_0)+F(\varphi),
\end{eqnarray*}
where $F=F_3-F_2$.
We need an independent equation for $\lambda$, so we
define the scalar product $(f,g)=\int \dd x f(x)^*g(x)$
and we normalize $\psi_0$ so that $(\psi_0,\psi_0)=1$.
Then we use the fact that, for any $g$, $(\psi_0,Mg)=0$ to write
\begin{eqnarray*}
-\lambda \big(\psi_0,\frac{\delta F_1(\varphi_0)}{\delta \psi_0}\big)=
(\psi_0,F_1(\varphi_0))+(\psi_0,F(\varphi)).
\end{eqnarray*}
Let $a=-\big(\psi_0,{\delta F_1(\varphi_0)}/{\delta \psi_0}\big)$,
we find
\begin{eqnarray*}
\lambda =
(\psi_0,F_1(\varphi_0))/a+(\psi_0,F(\varphi))/a.
\end{eqnarray*}
To obtain the second equation, we define $\Lambda$ as
the projector onto $K^{\perp}$, and we assume
that $\Lambda {\delta F_1(\varphi_0)}/{\delta \psi_0}=0$.
Thus we obtain
\begin{eqnarray*}
M\phiperp=\Lambda \big( F_1(\varphi_0)+F(\varphi)\big).
\end{eqnarray*}
If $G_0$ was the original Green function for $M$,
we define $G=\Lambda G_0 \Lambda$ and we obtain
\begin{eqnarray*}
\phiperp=G\big( F_1(\varphi_0)+F(\varphi)\big).
\end{eqnarray*}
We can group these two equations into a single one
\begin{eqnarray}
\lambda &=&
(\psi_0,F_1(\varphi_0))/a+(\psi_0,F(\varphi_0+\phiperp+\lambda\psi_0))/a
\nonumber\\
\phiperp &=& G\big( F_1(\varphi_0))+
G\big( F(\varphi_0+\phiperp+\lambda\psi_0)\big). \label{eqphiperp}
\end{eqnarray}

This equation is now in the form required for the application
of Butcher's method which writes the perturbative solution as
\begin{eqnarray*}
\lambda &=& \Phi^1_0+\sum_t \frac{1}{\sigma(t)} \Phi^1(t),\\
\phiperp(x) &=& \Phi^2_0(x)+\sum_t \frac{1}{\sigma(t)} \Phi^2(t;x).
\end{eqnarray*}
The zero-order terms are
\begin{eqnarray*}
\Phi^1_0  &=& (\psi_0,F_1(\varphi_0))/a,
\\
\Phi^2_0(x) &=& \int \dd y G(x,y)
 F_1(\varphi_0(y)).
\end{eqnarray*}
For the roots, the functions $\Phi^1$ and $\Phi^2$ are
defined by
\begin{eqnarray*}
\Phi^1(\tun)  &=& (\psi_0,F(\varphi_0))/a,
\\
\Phi^2(\tun;x) &=& \int \dd y G(x,y) F(\varphi_0(y)).
\end{eqnarray*}
For a tree $t=B_+(t_1,\dots,t_k)$, they are defined 
recursively by 
\begin{eqnarray*}
\Phi^1(t)  &=& \sum_{j_1\dots j_k}(1/a) \int \dd y
  \psi_0(y)\frac{\delta^k F(\varphi_0(y))}
  {\delta \Phi^{j_1}\cdots \delta \Phi^{j_k}} \\&&\times
  \Phi^{j_1}(t_1)\dots\Phi^{j_k}(t_k)\\
\Phi^2(t;x)  &=&  \sum_{j_1\dots j_k}\int \dd y G(x,y)
  \frac{\delta^k F(\varphi_0(y))} 
  {\delta \Phi^{j_1}\cdots \delta \Phi^{j_k}} \\&&\times
  \Phi^{j_1}(t_1)\dots\Phi^{j_k}(t_k).
\end{eqnarray*}
In the last expression, for $j_i=1$, then
$\delta/\delta\Phi^{j_i}=\delta/\delta\psi_0(y)$
$\Phi^{j_i}(t_i)=\Phi^1(t_i)$, and for $j_i=2$, then
$\delta/\delta\Phi^{j_i}=\delta/\delta\Phi^2(z_i)$,
$\Phi^{j_i}(t_i)=\Phi^2(t_i;z_i)$, and an integral
is implicitly assumed over the variable $z_i$.

This was just an example of the general strategy
available for the treatment of zero modes.
In specific problems, it might be more efficient to 
take also account of the dependence of $F_0(\varphi)$
on $\lambda$.

As a last point, we want to show on a simple example, that
using a modified Green function can make the problem
much better behaved. 
The example we want to discuss is
\begin{eqnarray*}
F_0(\varphi(x))= - \frac{\dd^2 \varphi(x)}{\dd x^2}
 +\varphi(x)-2g\varphi(x).
\end{eqnarray*}
This problem as a nontrivial solution
\begin{eqnarray*}
\varphi_0(x)=\frac{1}{\sqrt{g}} \frac{1}{\cosh x}.
\end{eqnarray*}
The linearization gives
\begin{eqnarray*}
M(x,y)=\frac{\delta F_0(\varphi_0(x))}{\delta \varphi(y)}=
(- \frac{\dd^2 }{\dd x^2}+1 -\frac{6}{\cosh^2 x})\delta(x-y).
\end{eqnarray*}
The equation $M\psi=0$ has two solutions: a normalizable
solution $\psi_0(x)$ and a  non normalizable one $\psi_1(x)$.
\begin{eqnarray*}
\psi_0(x) &=& \frac{\sinh x}{\cosh^2 x}, \\
\psi_1(x) &=& 3 \frac{x \sinh x - \cosh x}{\cosh^2 x} + \cosh x. \\
\end{eqnarray*}

The kernel of $M$ is the one dimensional subspace generated
by $\psi_0(x)$. By Wronski's method (\cite{Arfken},
p. 900), the Green function for $M$ is
\begin{eqnarray*}
G^0(x,y)= -\frac{1}{2} \psi_0(x_<) \psi_1(x_>),
\end{eqnarray*}
where, $(x_<,x_>)=(x,y)$ if $x<y$, and
$(x_<,x_>)=(y,x)$ if $x>y$. However, this Green function is
not well behaved because of the term $\cosh x$ which
gives divergent integrals. If calculations are made
with $G^0(x,y)$, the apparatus of renormalization theory
has to be used. This can be avoided by defining a 
well-behaved function $G^1=G^0+\delta G$, where
$\delta G(x,y)= \psi_0(x_<) \cosh(x_>)/2$. It can be
checked that $G^1(x,y)$ is now exponentially decreasing.
The equation $\Lambda G^0 M \phiperp=\phiperp$ becomes
\begin{eqnarray*}
\Lambda G^1 M \phiperp = \phiperp - \Lambda \delta G M \phiperp.
\end{eqnarray*}
Therefore, Eq.(\ref{eqphiperp}) becomes
\begin{eqnarray*}
\phiperp &=& \Lambda G^1\Lambda \big( F_1(\varphi_0)+
F(\varphi_0+\phiperp+\lambda\psi_0)\big)\\&&
-\Lambda \delta G M \Lambda \phiperp.
\end{eqnarray*}

Note that, in the sense of distributions
\begin{eqnarray*}
\int \dd z \delta G(x,z)M(z,y)
&=& \left( -1+\frac{3}{2\cosh^2x}\right)\delta(x-y)\\&&
-\frac{3}{\cosh x}\theta(x-y)
\end{eqnarray*}
is no longer divergent when integrated over x.

A method was proposed to write the general term of the
perturbative solution of a nonlinear problem.
This method is a nonlinear generalization of the Born
expansion. Although Butcher series are not widely known
in Physics, they are commonly used in Numerical Analysis,
where they have proved their power \cite{Butcher,Hairer},
and where many results are now available.

From the theoretical point of view,
Butcher series enable us to {\sl write} the general
term of the perturbative solution of nonlinear equations.
And since we can write it, we can manipulate it to
investigate stability,
long-time behavior, convergence, bifurcation, transition to chaos,
resummation, and so on. 

From the practical point of view,
the perturbation of a soliton is usually calculated by solving 
recursive differential equations, and the computation
becomes very cumbersome beyond the first order
\cite{Hasegawa}. By contrast, Butcher series offer a systematic
method which is easily implemented on a computer.

In Ref.\cite{Hasegawa}, several perturbations of the
optical soliton are reviewed. All can be considered
as induced by a change of variable. Such a change of variable
gives a modified nonlinear equation with a modified 
linearization $M'$. The Butcher series solution 
gives the general term of the perturbative solution for
this modified problem. Moreover, the non-linear change
of variable can be also implemented with trees.\cite{Wright} 
This, together with the fact that rooted trees have
recently appeared 
in singularity theory \cite{Kunkel}, renormalization
of quantum fields \cite{Kreimer}, and non-commutative
geometry \cite{ConnesM} 
indicate that they provide us with a tool particularly
well adapted to nonlinear problems.


%
%

%
%

\end{document}